**Abstract:** We show how the results given by several authors relatively to the mass of a density peak are changed when small scale substructure induced by dynamical friction are taken into account. The peak mass obtained is compared to the result of Peacock & Heavens (1990) and to the peak mass when dynamical friction is absent to show how these effects conspire to reduce the mass accreted by the peak.

## 1. Introduction

The origin and evolution of large scale structure is today the outstanding problem in Cosmology. In this context, a remarkable role has been played by the standard Cold Dark Matter (CDM) model.

The main features of the model are a scale invariant spectrum of density perturbations, $P(k)$, of the type Harryson-Zeldovich, growing under gravitational instability and the supposition that the bulk of the material in the Universe should be non baryonic but under the form of Weakly Interacting Massive Particles (WIMPS) (Efstathiou 1990, Kolb & Turner 1990). The only addition to the model, required by observation, is the question of biasing, i.e. the fact that the distribution of group and clusters of galaxies is different from the distribution of cosmic material on large scale, the first being more clustered than the second. The physical origin of such a biasing is not yet totally clear even though several mechanisms have been proposed (Rees 1985, Dekel & Rees 1987, Colafrancesco et al.1995).

In biased galaxy formation theory, it is assumed that structures of size $R_f$ form around the local maxima of the density field, smoothed on the filtering scale $R_f$. The linear density perturbations evolve towards the non linear regime because of gravitational instability and collapse when the average density is, $\overline{\delta} \simeq 1$.

A peak of the primordial density fluctuation field has a probability to form a protostructure proportional to its central height:

$$\nu = \frac{\overline{\delta}(0)}{\sigma_0(R_f)} \tag{1}$$

where $\overline{\delta}(r)$ is the overdensity in the radius $r$ and $\sigma_0(R_f)$ the mass variance windowed by a filter of scale $R_f$. It is assumed that only peaks with $\nu > \nu_c = \frac{\delta_c}{\sigma_0(R_f)} = \frac{1.68}{\sigma_0(R_f)}$ form structures.

The assignement of a mass, $M_{pk}$, to a peak of the filtered density perturbation field, is a

Heavens 1990; Lucchin & Matarrese 1988; Ryden 1988; Colafrancesco et al. 1989) have given different solutions.

The first and simplest is to assign to the peak the mass obtained from the volume of the filter function, $R_f$:

$$M_{pk} = (2\pi)^{3/2} \rho_c R_f^3 \qquad (2)$$

(Efstathiou & Rees 1988) where $\rho_c$ is the critical density. This means that the mass contained within a given volume is produced by the filtering process and we suppose that an object of lenght $R_f$ collapses when the mean overdensity in $R_f$ is, as previously written, $\overline{\delta}_c \simeq 1$.

If otherwise we use a spherical filter, the mass is given by:

$$M_{pk} = \frac{4\pi}{3} R_f^3 \rho_c \qquad (3)$$

An alternative to those given is to estimate $M_{pk}$ by modelling a peak as a triaxial ellipsoid (Peacock & Heavens 1990; Colafrancesco et al. 1989) obtaining a distribution of mass for a given $\nu$:

$$M_{pk} = \frac{4\pi}{3} \rho_c a_1 a_2 a_3 \qquad (4)$$

where $a_1$, $a_2$, $a_3$ are the semiaxes of the ellipsoid. In terms of $\nu$, $\nu_c$, the parameter of ellipticity e, the parameter of oblateness p, the local curvature parameter of the peak and the mass $M_* = \frac{4\pi}{3} R_*^3 \rho_c$, where $R_* \propto R_f$ we have:

$$M_{pk}(\nu, \nu_c, M; e, p, x) = (1 - f_{pk}\frac{\nu}{\nu_c})^{3/2} (\frac{2\nu}{\gamma})^{3/2} \left\{ [(1+p)^2 - 9e^2](1-2p)x^3 \right\}^{-1/2} M_* \qquad (5)$$

(Colafrancesco et al. 1989), where $f_{pk}$ is a term simulating the mass loss by the peak due to tidal interactions with neighbours, $\gamma$ is a parameter that can be expressed in terms of the spectral moments (see Bardeen et al 1986):

$$\sigma_j^2 = \int \frac{k^2 dk}{2\pi^2} P(k,t) k^{2j} \qquad (6)$$

Integrating Eq. (5) over e and p we have:

$$M_{pk} = \left(1 - f_{pk}\frac{\nu}{\nu_c}\right)^{3/2} \mu(\gamma, \nu\gamma) M_* \qquad 0 \leq \gamma\nu \leq 1.5 \qquad (7)$$

where $\mu(\gamma, \gamma\nu)$ is a function given by Colafrancesco et al.(1989) (Eq. 6).

Peacock & Heavens (1990) give another expression for the peak mass:

$$M_{pk} = \frac{2^{3/2}(4\pi)\rho_c R_*^3}{\gamma^3 + (\frac{0.9}{\nu})^{3/2}} \qquad 0.5 \leq \gamma \leq 0.8 \qquad (8)$$

The definition of the peak mass previously discussed strictly depends on the filtering process of the density field and on the filtering scale, which in turn depends on the physical process which set the threshold $\nu_c$. The exact analytic form of the filtering function is not known and it is often taken to be a Gaussian. At the same time there is not a precise relation between the filtering scale, $R_f$, and the mass of a class of objects (as we have seen in the Eq. (2), Eq. (3) and following). A more complex method, trying to overcome these problems, is that of Appel & Jones (1990). It is based on the observation that, during the process of filtering of the density field when $R_f$ is increased, some regions lying over the selected threshold grow in size while others decrease. The ones that decrease in size will vanish for a certain value of $R_f$. The point of vanishing is the position of the local maximum and $R_f$ is its radius. In this way the mass of a peak is chosen by the same procedure of filtering.

Finally, there is another procedure to define a peak mass and it is connected to the definition of a binding radius for the structure (Hoffmann & Shaham 1985). It is directly connected to the process of collapse of a density perturbation.

As shown by Antonuccio-Delogu & Colafrancesco (1994) the evolution of structures like Clusters of galaxies is influenced by the dynamical friction induced by the substructure present in them in accordance with CDM model. Consequently the binding radius and the mass of a density peak must be influenced by dynamical friction. These observations lead us to recalculate the peak mass taking into account this dissipative effect.

In this paper we use the definition of mass peak given by Hoffmann & Shaham (1985) to calculate the peak mass and we show how it must be changed due to the presence of substructure in protostructure, inducing dynamical friction. Finally, we compare this result to that given by Peacock & Heavens (1990).

## 2. Mass peak from binding radius

There are at least two ways to assign a radius to a peak of density: a statistical one, (Ryden 1988), and a dynamical one, (Hoffmann & Shaham 1985).

In the first criterion the binding radius of the region, $r_b$, is given by the solution of the equation:

$$<\overline{\delta}(r)> = <(\overline{\delta}(r) - <\overline{\delta}(r)>)^2>^{1/2} \tag{9}$$

At radius $r \ll r_b$ the motion of particles is predominantly toward the peak while when $r \gg r_b$ particles are not bound to the peak.

shell collapses in a time, $T_{c0}$, smaller than the age of the universe $t_0$:

$$T_{c0}(r) \leq t_0 \tag{10}$$

This last criterion, differently from the previous one, contains a prescription particularly connected with the physics of the collapse process of a shell. It supposes that only the regions that have collapse time less than a Hubble time form structure. The time of

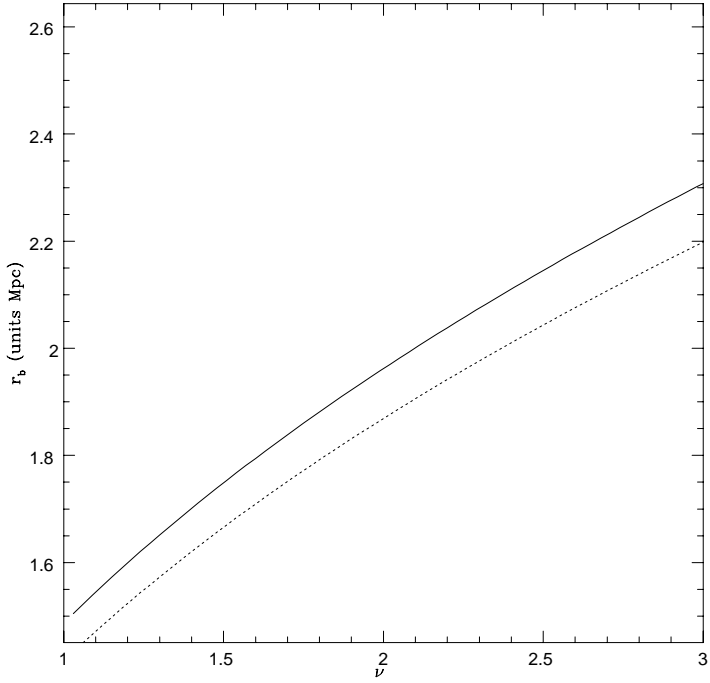

**Fig. 1.** Variation of the binding radius $r_b$ with $\nu$. The solid line is the binding radius in absence of dynamical friction, while the dashed line is the same as in presence of dynamical friction. The filtering radius used is, $R_f = 1 \, Mpc$

collapse, $T_{c0}(r)$, can be obtained using the equation:

$$T_{c0}(r) = \frac{3\pi}{2} \frac{(1+\overline{\delta}(r))}{\overline{\delta}(r)^{3/2}} \tag{11}$$

Gunn & Gott (1972).
To obtain the average density $\overline{\delta}$ we used the equation given by Bardeen et al. (1986):

$$\delta(r) = A \left\{ \frac{\nu \xi(r)}{\xi(0)^{1/2}} - \frac{\theta(\nu\gamma, \gamma)}{\gamma \xi(0)^{1/2}(1-\gamma^2)} \left[ \gamma^2 \xi(r) + \frac{R_*^2 \nabla^2 \xi(r)}{3} \right] \right\} \tag{12}$$

$\xi(r)$ is the correlation function of two points, $\gamma$ and $R_*$ two constants obtainable from the spectrum (see Bardeen et al. 1986) and finally $\theta(\gamma\nu,\gamma)$ is a function given in the quoted paper (Eq. 6.14). The average density, $\overline{\delta}$, inside the radius $r$ in a spherical perturbation is given by:

$$\overline{\delta} = \frac{3}{r^3} \int_0^r dx \delta(x) x^2 \qquad (13)$$

We calculated the time of collapse, $T_{c0}(r)$, using Eq. (11) with the density profile given in Eq. (12). We repeated the calculation of $T_{c0}(r)$ for $1.5 < \nu < 3$ and we applied the condition given in Eq. 10 to the curves $T_{c0}(r)$ previously obtained. The result is the plot in Fig. 1 for the binding radius $r_b$ versus $\nu$. Using the plot $r_b - \nu$ and the equation:

$$M_{pk} = \frac{4\pi}{3} r_b^3 \rho_c \qquad (14)$$

finally we obtained the peak mass. As expected, it is an increasing function of $\nu$.

## 3. Mass of a peak in presence of dynamical friction

The calculation of the peak mass, given in the Sect. 2, does not take into account the substructure present into a collapsed perturbation. In presence of it the collapse of a structure is modified (Antonuccio-Delogu & Colafrancesco 1994) and, as a consequence, the mass accreted by the structure is modified too.

The time of collapse in presence of dynamical friction is given by:

$$T_c(r) = T_{c0} \left\{ 1 - \frac{\lambda_0}{1 - \frac{2^{1/2} \pi [1+\overline{\delta}(r)]^{3/2}}{3c(\overline{\delta})\overline{\delta}(r)^{3/2}}} \right\} \qquad (15)$$

where $\lambda_0$ is a function of the number of peaks and of its average mass (Eq. 23 in Antonuccio-Delogu & Colafrancesco 1994), while $c(\overline{\delta})$ is a function of the overdensity $\overline{\delta}$ (Eq. 27 in the quoted paper).

We calculated the collapse time $T_c(r)$ for a given value of $\nu$ using Eq. (15) and the density profile Eq. (12). We repeated the calculation procedure made in the previous section to determine the binding radius, $r_b$ for $1.5 \leq \nu \leq 3$. Using the plot $r_b$-$\nu$ and the Eq. (14) we finally obtained the peak mass, $M_{pk}$ in presence of dynamical friction. In Fig. 2 we compared this result with the peak mass in absence of dynamical friction and with the equation of Peacock & Heavens (1990).

The peak mass in absence of dynamical friction is comparable to that given by Peacock & Heavens (1990) while, as expected, dynamical friction reduces the peak mass with respect to systems in which it is absent. The definition of the peak mass is influenced by the presence of substructure and then from dynamical friction. The total mass density associated with peaks is consequently changed with also the construction of the mass function following the Press & Schechter (1974) prescription.

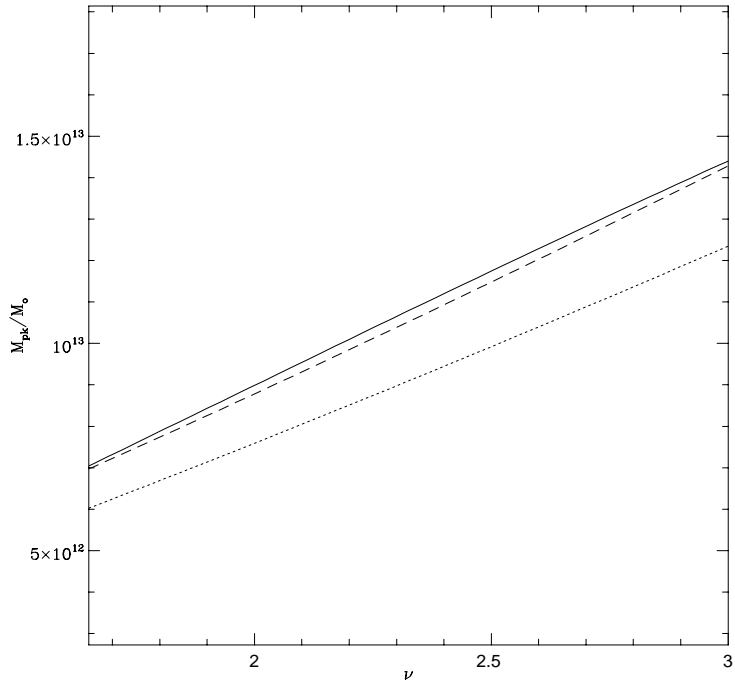

**Fig. 2.** Variation of the peaks mass $M_{pk}$ with $\nu$. The solid line representes the curve derived by Peacok & Heavens given in text; the dashed line rappresentes $M_{pk}$ in absence of dynamical friction and the dotted line rappresentes $M_{pk}$ in presence of dynamical friction

This analysis shows that dynamical friction reduces the mass bound by a peak with respect to that they should acquire if this dissipative effect was not present.

*Acknowledgements.* We are grateful to V. Antonuccio-Delogu for some interesting discussion and for several useful comments. A particular thanks to Antonio Magazzú for his comments on the paper.

### References

Antonuccio-Delogu, V., Colafrancesco, S., 1994, Ap.J. 427, 72

Appel, L., Jones, B.J.T., 1990, MNRAS 245, 522

Bardeen, J.M, Bond, J.R., Kaiser, N., Szalay, A.S., 1986, Ap.J 304, 15

Colafrancesco, S., Lucchin, F., Matarrese, S., 1989, Ap.J 345, 3

Colafrancesco, S., Antonuccio, V., Del Popolo, A., 1995, Ap. J

Dekel, A., Rees, M., 1987, Nature 326, 455

Efstathiou, G., Rees, M.J., 1988, MNRAS, 230, 5p.

Efstathiou, G., 1990, in "The physics of the early Universe", eds Heavens, A., Peacock, J., Davies, A., (SUSSP)

Gunn, J.E., Gott, J.R., 1972, Ap.J 176, 1

Hoffmann, Y., Shaham, J., 1985, Ap.J. 297, 16.

Kolb, E.W., Turner, M.S., 1990, The early Universe (Addison-Wesley)

Lucchin, F., Matarrese, S., 1988, Ap.J 330, 21

Peacok, J.A., Heavens, A.F., 1990, MNRAS 243, 133

Press, W.H, Schechter, P.L, 1974, Ap.J 187, 425

Rees, M.J., 1985, MNRAS, 213, 75p

Ryden, B.S., 1988b, Ap.J 333, 78